\documentclass[
aip,
jap,
preprint,
preprintnumbers,
showpacs,
amsmath,
amssymb]{revtex4-1}
\usepackage{graphicx}
\usepackage{dcolumn}
\usepackage{bm}
\usepackage{textcomp}
\usepackage{revsymb}
\usepackage{amssymb}

\begin{document}


\title{Nucleation of Ge clusters  at high temperatures on Ge/Si(001) wetting layer}

\author{Larisa V. Arapkina}
\email{arapkina@kapella.gpi.ru}
\noaffiliation

\author{Vladimir A. Yuryev}
\email{vyuryev@kapella.gpi.ru} 
\homepage{http://www.gpi.ru/eng/staff\_s.php?eng=1\&id=125}
\noaffiliation

\affiliation{A.\,M.\,Prokhorov General Physics Institute of the Russian Academy of Sciences, 38 Vavilov Street, Moscow, 119991, Russia}

\date{\today}%

\begin{abstract}
 Difference in nucleation of Ge quantum dots   during Ge deposition at  low ($<$\,600{\textcelsius}) and high ($\agt$\,600{\textcelsius}) temperatures on the Si(001) surface is studied by high resolution scanning tunneling microscopy and {\it in situ} reflected high-energy electron diffraction. Two process resulting in appearance of $\{105\}$-faceted clusters on the Ge wetting layer have been observed at high temperatures: Pyramids have been observed to nucleate via the previously described formation of strictly determined structures, resembling blossoms, composed by 16 dimers grouped  in pairs and chains of 4 dimers on tops of the wetting layer $M\times N$ patches, each on top of a separate single patch, just like it goes on at low temperatures; an alternative process consists in faceting of shapeless heaps of excess Ge atoms which arise in the vicinity of strong sinks of adatoms, such as pits or steps. The latter process is not observed at low temperatures; it is typical only for the high-temperature deposition mode.
 \end{abstract}



\maketitle

\section{Introduction }

Issues of  nucleation of  $\{105\}$-faceted Ge clusters  under different growth conditions on Si(001) have attracted an attention of researches since the discovery of huts \cite{Mo} (see, e.g., Refs.\,\onlinecite{Iwawaki_initial,Nucleation,Goldfarb_2005,Vailionis,Jesson_Kastner_Voigt} and a brief review  presented in the introduction to Ref.\,\onlinecite{classification}). In recent articles, \cite{Hut_nucleation,CMOS-compatible-EMRS,initial_phase} we described a process of hut nucleation at low ($<$\,600{\textcelsius}) temperatures of Ge deposition consisted in formation of strictly determined structures composed by 16 dimers grouped  in pairs and chains of 4 dimes on tops of the wetting layer $M\times N$ patches, each on top of a separate single patch (Fig.\,\ref{fig:nuclei360}). However, it has been unclear thus far if the same nucleation process takes place at high temperatures ($\agt$\,600{\textcelsius}) or there are some different cluster formation mechanisms peculiar for this growth mode. Previous articles by different authors \cite{Nucleation,Goldfarb_2005}, who had explored hut formation in the process of the gas-source molecular beam epitaxy (MBE) based on decomposition of germane (GeH$_4$) at high temperatures on a Si surface,  indicated that huts arose near  wetting layer (WL) irregularities, such as  $\{105\}$-faceted pits, which appeared in thick hydrogen-rich WL prior to cluster nucleation, or [100] steps. In fact, it has been almost commonly adopted that huts always nucleate as a result of the latter process regardless the epitaxy technique applied for Ge film deposition; [100] steps of much thinner WL, which forms in the ultra-high vacuum (UHV), have been assumed to be the required irregularities in the case of UHV MBE. 

Being aware of a different hut nucleation mechanism at low temperatures we have decided to re-examine the outset of hut array formation at high temperatures to verify or refute the commonly adopted opinion. The results of our investigations are presented in the current article.

  \section{Equipment and Experimental Techniques }    

The experiments were carried out using an integrated UHV
instrument 
\cite{[{See, e.\,g., }]classification,stm-rheed-EMRS,CMOS-compatible-EMRS,VCIAN2011} built on the basis of the Riber~SSC\,2
surface science center with the  EVA\,32 MBE chamber   connected through a gate to the
STM GPI-300 UHV scanning tunneling microscope
\cite{STM_GPI-Proc,*gpi300}.
A preliminary
annealing and outgassing chamber, as well as facility for reflected high energy electron diffraction (RHEED) analysis and a number of complimentary techniques of surface exploration, are also available in the instrument.\cite{VCIAN2011}

The pressure of about $5\times 10^{-9}$\,Torr was kept in the preliminary
annealing chamber. 
The MBE chamber was evacuated down to about $10^{-11}$\,Torr before processes; the pressure grew to nearly $2\times 10^{-9}$\,Torr at most during the sample surface deoxidization process \cite{stm-rheed-EMRS,our_Si(001)_en,*phase_transition} and $10^{-9}$\,Torr during Ge or Si deposition.   
The residual gas pressure did not exceed $10^{-10}$\,Torr in the STM chamber.

Sources with the electron beam evaporation were used for Ge  deposition.
 The deposition rate and  coverage were measured using
the Inficon Leybold-Heraeus  XTC\,751-001-G1 film thickness monitor equipped with the graduated
in-advance quartz sensors installed in the MBE chamber. 
Tantalum radiators were used  for sample heating from the rear side in both
preliminary annealing and MBE chambers. The temperature was
monitored with chromel--alumel and tungsten--rhenium thermocouples of the heaters
in the preliminary annealing and MBE chambers, respectively. The
thermocouples were mounted in vacuum near the rear side of the
samples and {\it in situ} graduated beforehand against the
IMPAC~IS\,12-Si pyrometer which measured the sample temperature  through chamber windows. 
The  composition of residual atmosphere in the MBE
camber was monitored using the SRS~RGA-200 residual gas analyzer
before and during the process.

The STM tip was {\it ex situ} made of the tungsten
wire and cleaned by ion bombardment \cite{W-tip} in a special UHV
chamber connected to the STM one.

In this work, the images were obtained in the
constant tunneling current ($I_{\rm t}$) mode at the room
temperature. The STM tip was zero-biased while the sample was
positively or negatively biased ($U_{\rm s}$) when scanned in empty-states
or filled-states imaging mode.
Original firmware \cite{STM_GPI-Proc,*gpi300,STM_calibration} was
used for data acquisition; the STM images were processed afterward
using the WSxM software \cite{WSxM}.

RHEED analysis was performed {\it in situ} during sample annealing or Ge deposition by the Staib Instruments RH20 diffractometer.


Initial samples for STM were  $8\times 8$~mm$^{2}$ squares cut from the
specially treated commercial boron-doped    Czochralski-grown (CZ) Si$(100)$ wafers
($p$-type,  $\rho\,= 12~{\Omega}\,$cm).  After washing and chemical treatment
following the standard procedure described elsewhere
\cite{etching} (which included washing in
ethanol, etching in the mixture of HNO$_3$ and HF and rinsing in the
deionized water \cite{VCIAN2011,cleaning_handbook}), the silicon substrates were mounted on the
molybdenum STM holders and inflexibly clamped with the tantalum
fasteners; the STM holders were placed in the holders for MBE made
of molybdenum with tantalum inserts. 
Then, the substrates were outgassed at the temperature of
around {565{\textcelsius}} for more than 6\,h in the preliminary
annealing chamber. After that, the substrates were
moved  into the MBE chamber 
where they were subjected to 
final annealing with the maximum temperature of
about {925{\textcelsius}} \cite{classification,stm-rheed-EMRS,VCIAN2011,our_Si(001)_en,phase_transition}. Then, the temperature was rapidly
lowered to about {750{\textcelsius}}. The rate of the further cooling was
around 0.4{\textcelsius}/s that corresponded to the quenching mode
applied in Ref.\,\onlinecite{stm-rheed-EMRS}. The resultant Si(001)
surfaces were completely purified of the
oxide film and atomically clean \cite{classification,stm-rheed-EMRS,VCIAN2011,our_Si(001)_en,phase_transition}; their structure was investigated in detail in Refs.\,\onlinecite{stm-rheed-EMRS,our_Si(001)_en,phase_transition}.

Ge was deposited directly on the deoxidized Si(001) surface; the
deposition rate ($dh_{\rm Ge}/dt$) was varied from about $0.1$ to $0.15$\,\r{A}/s. The substrate
temperature during Ge deposition $(T_{\rm gr})$ was {360{\textcelsius}} for the low-temperature mode   and 600 or {650{\textcelsius}} for the  high-temperature mode.
The rate of the sample cooling down to the room temperature was approximately
0.4{\textcelsius}/s after the deposition.
After  cooling, the prepared samples were placed for
analysis into the STM chamber.


\section{Nucleation of germanium clusters  at low and high temperatures}

\subsection{Scanning Tunneling Microscopy}

\subsubsection{Low Temperature Mode}

\begin{figure*}
\includegraphics[scale=.8]{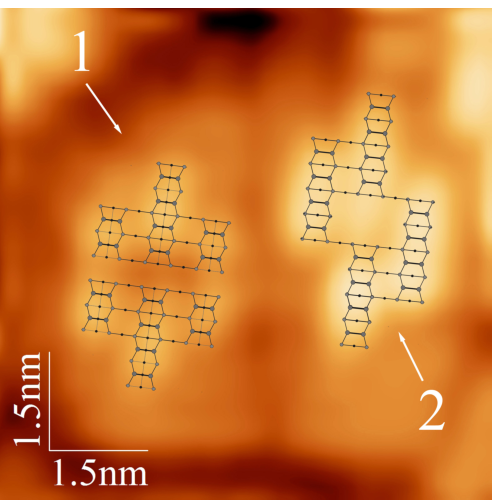}(a)
\includegraphics[scale=1]{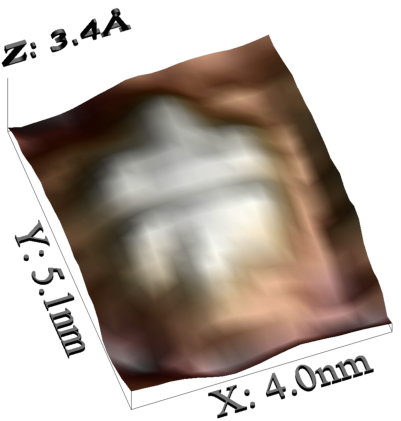}(b)
\includegraphics[scale=.8]{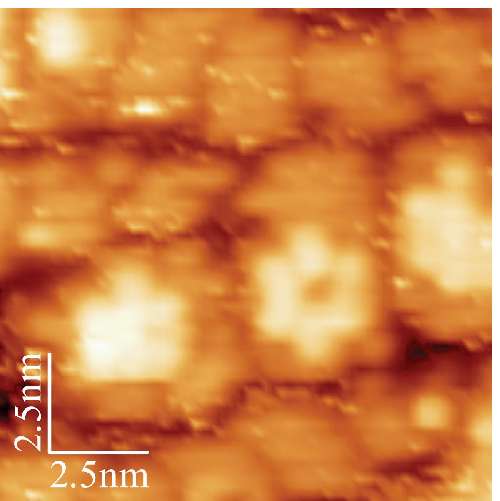}(c)
\caption{\label{fig:nuclei360}
(Color online)
STM empty-state images of hut nuclei on Ge wetting layer formed at $T_{\rm gr} =$ 360{\textcelsius}:
(a) pyramid (1) and wedge (2) nuclei  on the adjacent
$M \times N$ patches of wetting layer; $h_{\rm Ge}=$ 6\,\AA; the structural models \cite{Hut_nucleation,CMOS-compatible-EMRS} are superimposed on the corresponding image; (b) $h_{\rm Ge}=$ 6\,\AA;\cite{classification}
(c)  $h_{\rm Ge}=$ 5.4\,\AA.
}
\end{figure*}

 {\it Nucleation of Ge clusters at low temperatures} ($< 600$\,\textcelsius) was recently described by us in a number of articles.\cite{Hut_nucleation,CMOS-compatible-EMRS,VCIAN2011} We have demonstrated that huts form via parallel nucleation of two characteristic embryos different only in symmetry and composed by epitaxially oriented Ge dimer pairs and  chains of four dimers on tops of the wetting layer $M\times N$ patches: an individual embryo for each species of huts---pyramids or wedges (Fig.~\ref{fig:nuclei360}a--c). \cite{classification,CMOS-compatible-EMRS}
These nuclei always  arise on sufficiently large WL patches: there must be enough space for a nucleus on a single patch; a nucleus cannot form on more than one patch. \cite{Hut_nucleation,CMOS-compatible-EMRS} This fact may be explained in assumption of presence of the Ehrlich-Schwoebel barriers\cite{Schwoebel,*Ehrlich-Schwoebel} on sides of  WL patches which prevent the spread of dimer chais composing nuclei from one patch to another; nuclei form on bottoms of potential wells of the spatial potential relief associated with  the WL patches. As it follows from Ref.\,\onlinecite{initial_phase}, the total patch thickness (from Si/Ge interface to the patch top) rather than the mean thickness of WL controls the hut nucleation process. Further growth of a hut likely decreases and finally eliminates the potential barrier; then hut occupies adjacent WL patches.

The hut nuclei of both types arise at the same WL patch thickness of 4\,ML,\cite{CMOS-compatible-EMRS,initial_phase,VCIAN2011} at least at the initial phase of hut array formation, likely until growing huts start to contribute considerably to  consumption of migrating adatoms.\cite{classification}  Therefore, they form at the same WL stress to relieve it. Consequently, they appear at the same strain energy (and with equal probabilities ($P$),\cite{classification,Hut_nucleation}  hence, initially their free energy barriers ($W$) are the same magnitude since these quantities are related through the Boltzmann factor:\cite{Parallel_nucleation} $P \propto \exp\{-W/kT_{\rm gr}\}$).
This means that they are degenerate by the formation energy: 
if they had different formation energies they would first appear at different WL patch thicknesses.\cite{initial_phase} Their nucleation rates are also equal at the initial phase of array formation ([$dN_{\rm w}/dt]/[dN_{\rm p}/dt] = P_{\rm w}/P_{\rm p}$, w and p denote wedge and pyramid).

Parallel nucleation of two competitive formations (hut nuclei), different only in symmetry, on Ge WL is an amazing process. In Ref.\,\onlinecite{initial_phase}, we proposed a speculation which somewhat explains this phenomenon on the basis of modeling of Ge cluster formation energy performed in Ref.~\onlinecite{Domes_first}, according to which flat Ge islands likely occur on WL because of an energy benefit which arises in exposing the compressed \{105\} facets, rather than in relaxing the volumetric elastic energy. This speculation implies some difference in hut nucleation process from the usual the Stranski-Krastanow mechanism. However, this model certainly cannot explain the specific  symmetries of nuclei and the fact that only two nuclei shapes appear on WL patches. 

Another hypothesis might explain the latter facts. We reported in Ref.\,\onlinecite{initial_phase} the simultaneous formation of the $c(4\times 2)$ and $p(2\times 2)$ reconstructions on tops of adjacent WL patches; a pyramid nucleus was observed on the  surface of a $c(4\times 2)$-reconstructed patch.\cite{initial_phase,VCIAN2011} We suppose that WL patch top reconstruction determines the type of a hat which can form on a patch. If this is the case, the $c(4\times 2)$ reconstruction of the patch top enables  the nucleation of a pyramid, whereas the $p(2\times 2)$ reconstruction allows a wedge to arise on the patch. Domination of one of these reconstruction types may result in domination of a certain species of huts. 

Coexistence of the two reconstructions of WL patch tops seems to be more explicable. If they form simultaneously they have almost equal energies and correspond to two potential minima among possible variations of the $(2\times 1)$ dimer-row structure (they have two simplest possible symmetries). Let us consider two adjacent dimer rows forming on top of a WL patch. A fluctuation may give rise to parallel orientation of buckling of two neighbouring dimers from different rows (when one of the closest atoms from different dimer rows is in upper position and another is in lower position) or result in opposite orientation of dimer buckling (when both closest atoms from different rows are in upper or in lower position). The next dimers in each row will stand crosswise to the first one to minimise the energy; they will form the inverse configuration of pairs. 
Then the system should transit like a zipper to the state corresponding to one  of two possible potential minima: $p(2\times 2)$ for the case of the parallel buckling of initial dimers or  $c(4\times 2)$ for the opposite one. The resultant reconstruction will fill an entire top layer of the patch but it will not affect a state of the adjacent patch and any of the two possible reconstructions can arise on its top, that is observed experimentally.\cite{initial_phase} These reconstructions may be stable at low temperatures of Ge deposition and each species of huts may nucleate on a patch with a specific structure of its top.

Thus the, according to our assumption, fluctuations of relative buckling of only two dimers in adjacent  rows on top of each patch, which seem to be very likely, giving rise to  simultaneous parallel formation of two modifications of the $(2\times 1)$ structures with the simplest symmetries, may be the reason of parallel nucleation of the two known species of huts at low temperatures.

Remark that this hypothesis certainly needs a theoretical and experimental verification. A usual model of the Stranski-Krastanow hut cluster formation will also require modification if the hypothesis is true.

In conclusion of this section, we would like to consider an alternative scenario of  serial nucleation of huts which also may take place. Let us address again to Fig.~\ref{fig:nuclei360}a. One can see that the formation marked by the numeral 2 and considered as a wedge nucleus is composed by two couples of dimer rows, each row consists of four dimers (an arrow which starts at the numeral 2 indicates the lower right couple of rows). Such couples of 4-dimer sections of rows are often seen across the WL surface in the STM images.\cite{initial_phase,classification,CMOS-compatible-EMRS} They may be considered as cluster embryos  in the process of serial nucleation. Such embryo may compose a nucleus of a wedge  by pairing with another embryo and forming a structure marked by the numeral 2 in Fig.~\ref{fig:nuclei360}a. Then, after reconstruction, the latter structure may transform in the blossom-like formation considered as a nucleus of a pyramid (marked by 1 in Fig.~\ref{fig:nuclei360}a). Perhaps, this scenario gives a direction of thought to explain the decrease in the fraction of pyramids in the arrays which is observed as the Ge coverage increases.\cite{classification,CMOS-compatible-EMRS}
 
We note also that this scenario does not contradict the above hypothesis about the affect of the patch top reconstruction on hut nucleation process; both reasonings can  be easily combined to explain the whole process of hut nucleation.

\subsubsection{High Temperature Mode}

In contrast to the low-temperature mode, there are at least two ways of {\it nucleation of $\{$105$\}$-faceted Ge clusters at high temperatures}.\cite{VCIAN2011}

\begin{figure*}
\includegraphics[scale=1.055]{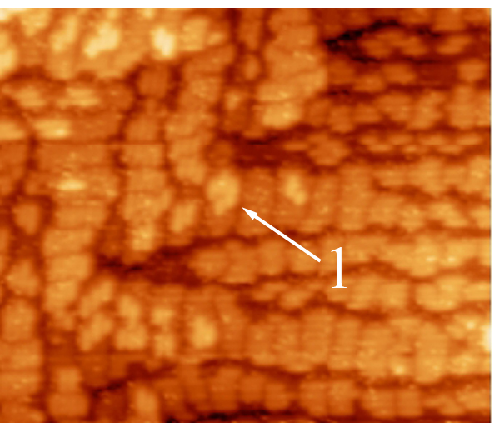}(a)
\includegraphics[scale=.5]{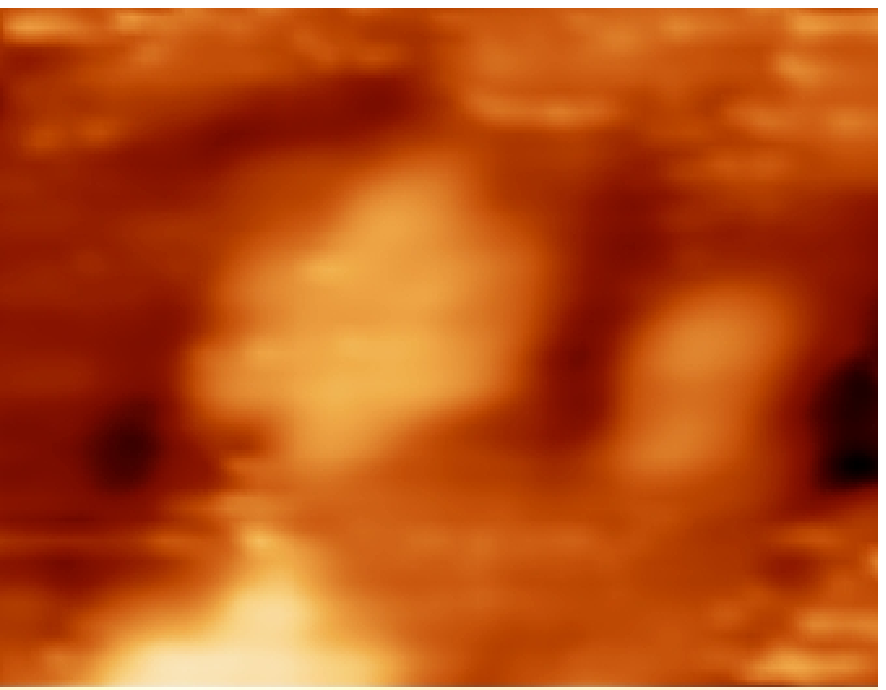}(b)\\
\includegraphics[scale=.55]{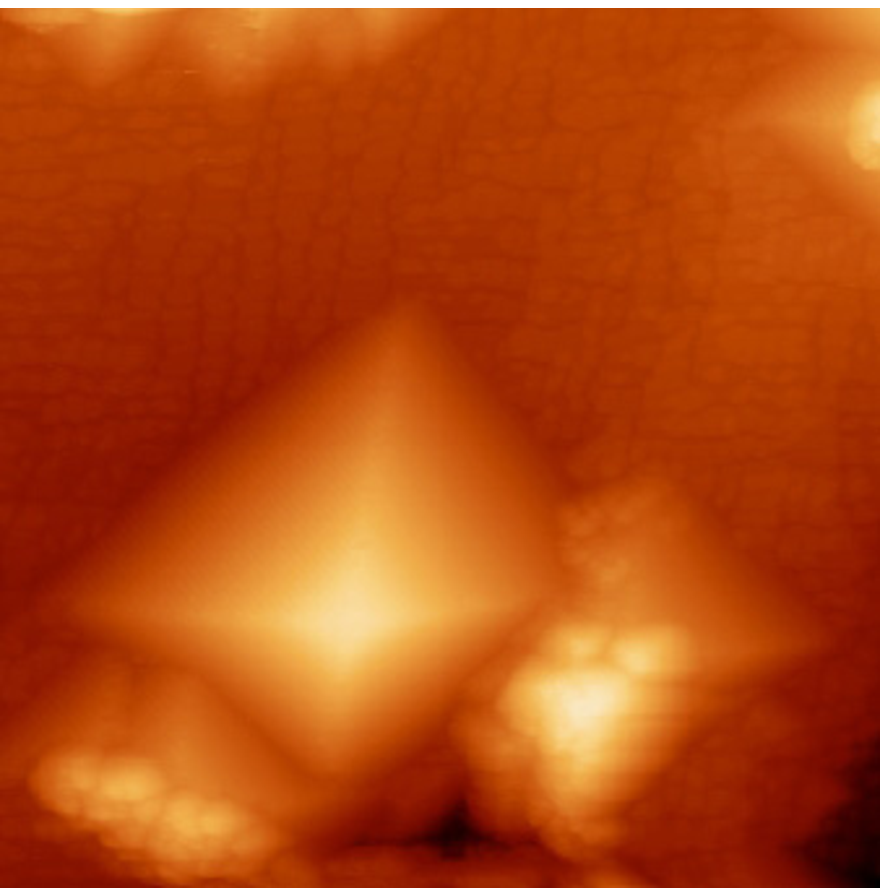}(c)
\includegraphics[scale=.55]{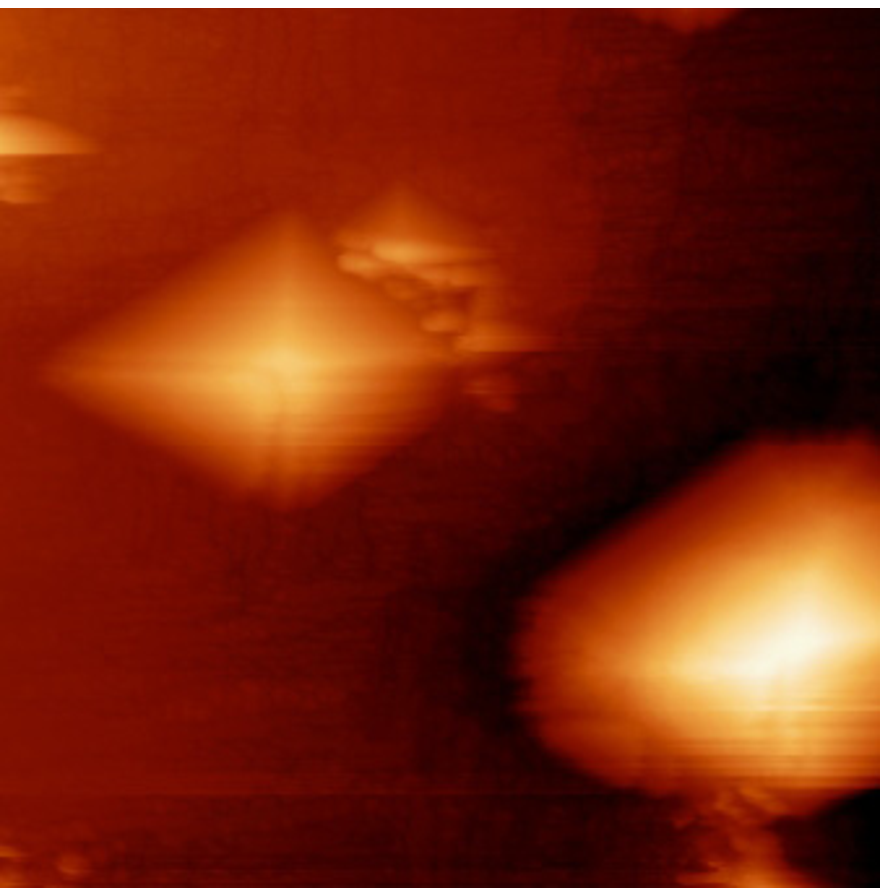}(d)
\caption{\label{fig:nuclei600}
(Color online)
STM empty-state images of hut nuclei (a, b) and pyramids (c, d) on Ge wetting layer  formed at high temperatures: 
(a, b) nuclei,  $h_{\rm Ge}=$ 5\,{\AA};
(a) $T_{\rm gr} =$ 600{\textcelsius}, $43\times 37$\,nm, (1) indicates the pyramid nucleus;
(b) $T_{\rm gr} =$ 650{\textcelsius},  $7.8\times 6$\,nm;
(c, d) pyramids, $h_{\rm Ge}=$ 6\,{\AA};
(c)  $T_{\rm gr} =$ 600{\textcelsius}, $110\times 110$\,nm;
(d) $T_{\rm gr} =$ 650{\textcelsius}, $200\times 200$\,nm.
}
\end{figure*}

 The first way is similar to the process of hut nucleation at low temperatures. Pyramids were observed to nucleate in such a way. Fig.~\ref{fig:nuclei600}a,b illustrates this statement: the pyramid nuclei, absolutely the same as those observed in the samples grown at low temperature, are seen on the WL patches in the images of the samples obtained at $T_{\rm gr} =600$ and  650{\textcelsius}, $h_{\rm Ge}$ = 5\,\r{A}. Their density was small, and they were mainly situated in the vicinity to large mature pyramids (Fig.~\ref{fig:nuclei600}c,d), which arise at early stages of Ge deposition and have much greater sizes (up to 8~nm in height at $h_{\rm Ge}$ = 6\,\r{A})\cite{Nanophysics-2011_Ge} than huts formed at low temperatures at the same values of $h_{\rm Ge}$.\cite{classification,CMOS-compatible-EMRS,VCIAN2011} The WL surface mainly consisted of  monoatomic steps and narrow terraces in these ares (Fig.~\ref{fig:nuclei600}b). Neither wedges nor wedge nuclei were observed. 

The absence of wedges or even wedge nuclei at high temperatures, in contrast to  low temperatures, waits for a  theoretical explanation. It does not contradict the proposed qualitative speculations, however.

\begin{figure*}
\includegraphics[scale=.42]{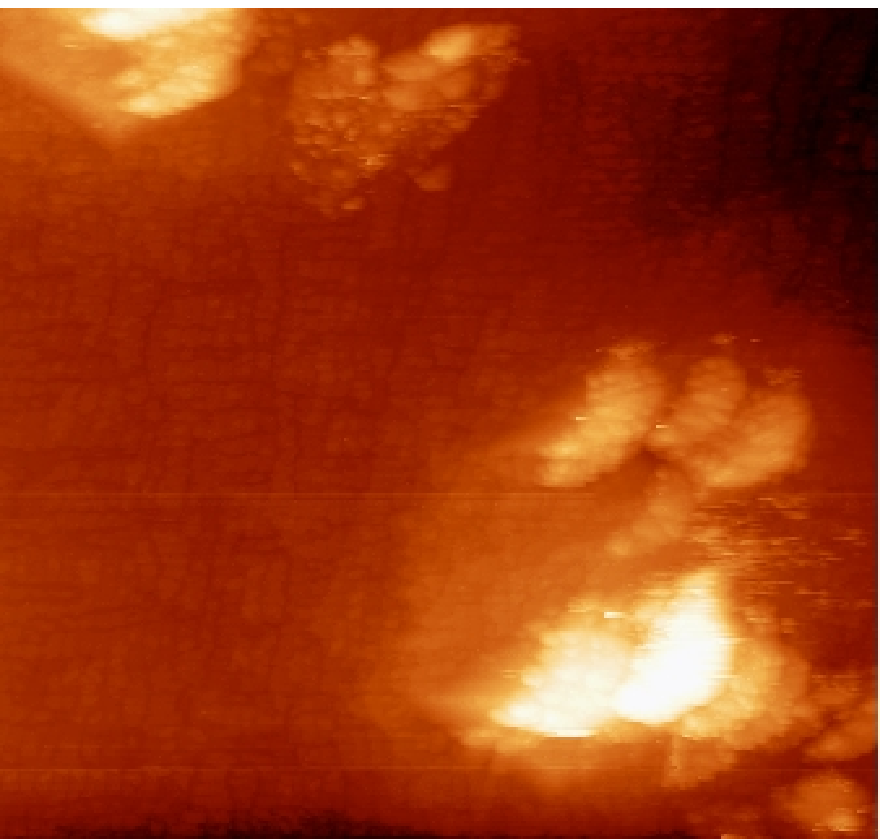}(a)
\includegraphics[scale=.4]{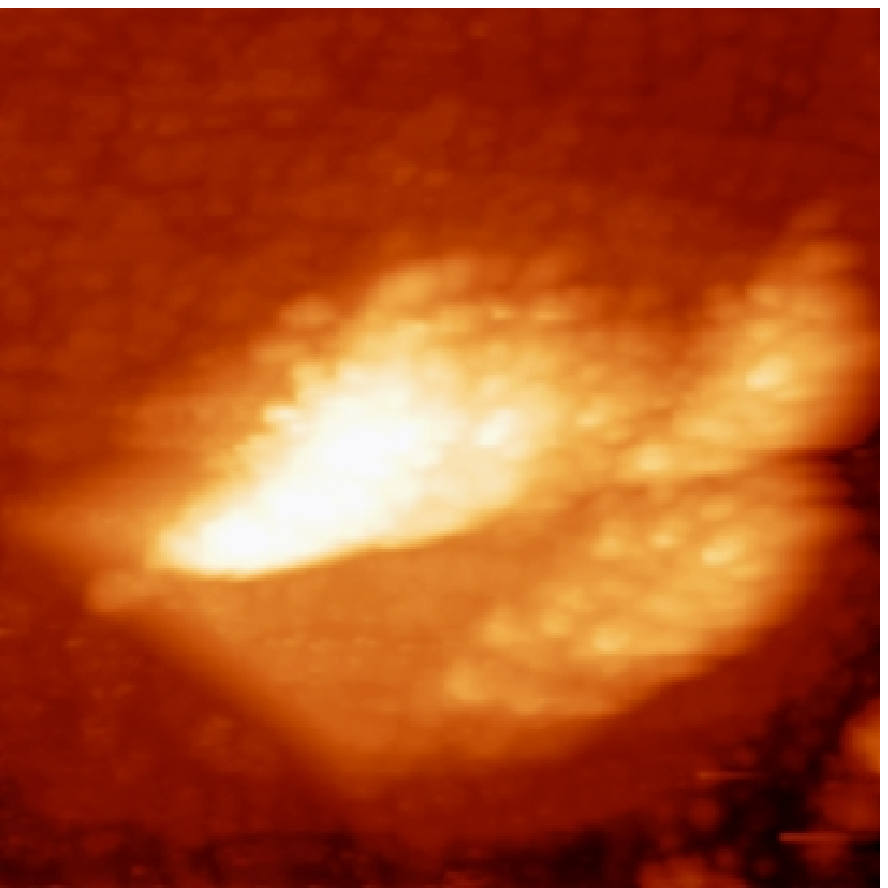}(b)\\
\includegraphics[scale=.4]{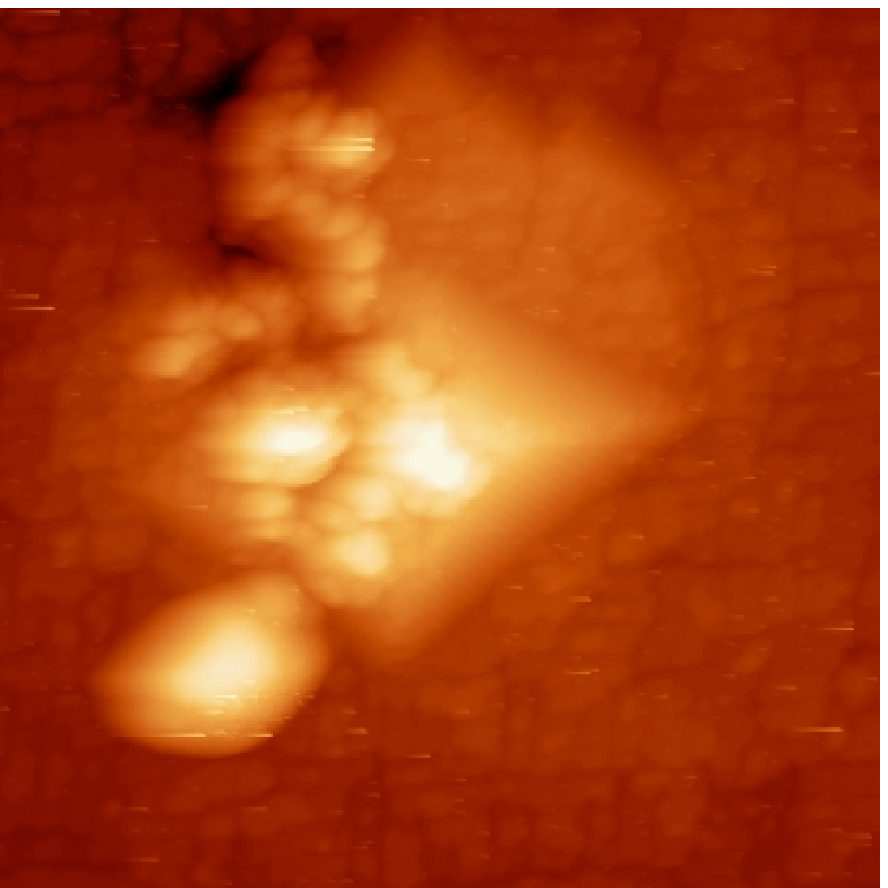}(c)
\includegraphics[scale=.4]{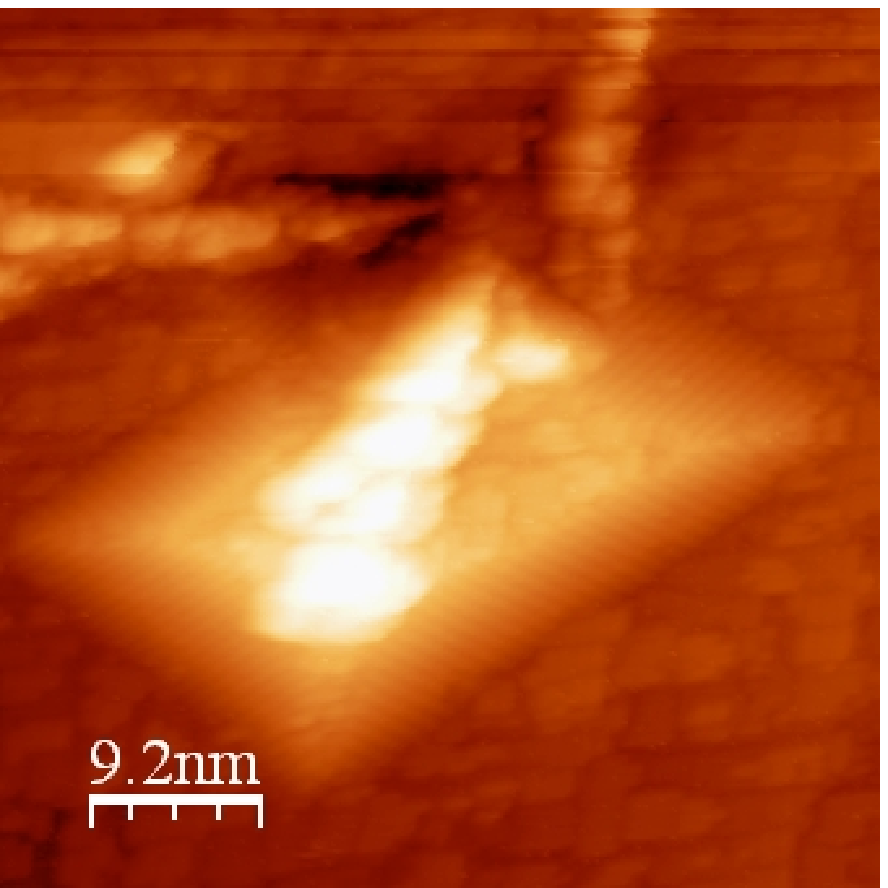}(d)
\includegraphics[scale=.4]{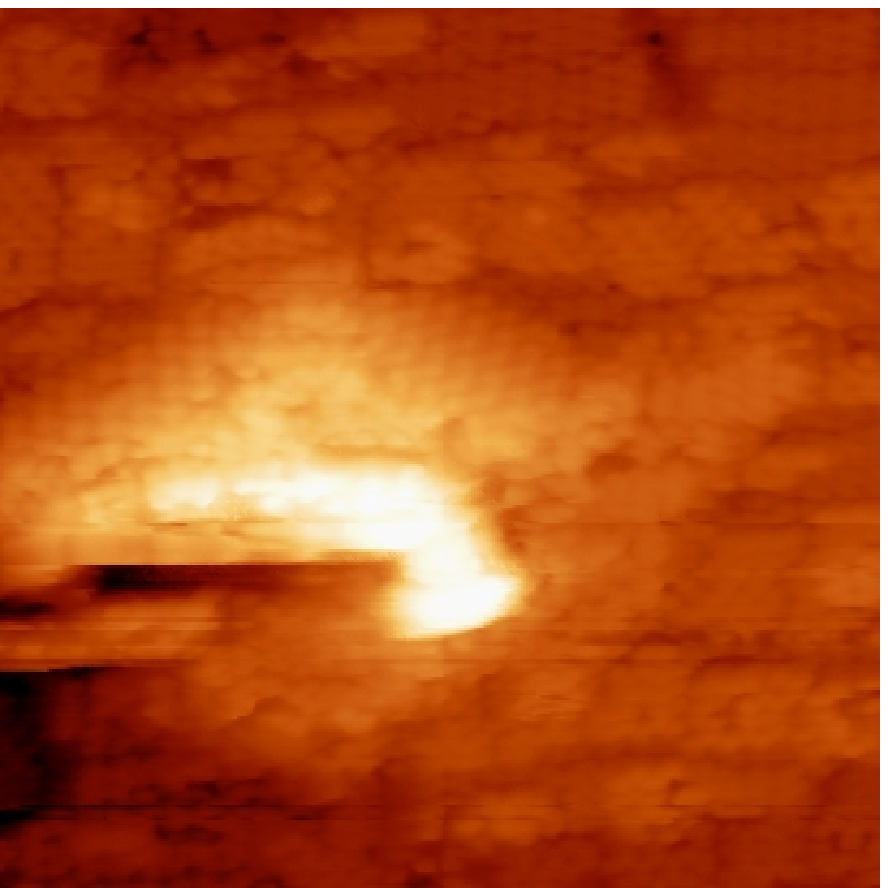}(e)
\caption{\label{fig:heap}
(Color online)
Formation of $\{$105$\}$ facets on shapeless heaps of Ge:
STM images of different phases of faceting, $T_{\rm gr} =$ 650{\textcelsius}, $h_{\rm Ge}=$ 5\,\AA;
(a) a shapeless Ge heap without faceting, $150\times 141$\,nm;
(b) the outset of faceting, $64\times 64$\,nm;
(c)--(e) effect of annealing at $T_{\rm gr} =$ 650{\textcelsius}:
developed $\{$105$\}$ facets are clearly seen;
(c) $72\times 72$\,nm;
(d) $46\times 46$\,nm;
(e) $23\times 23$\,nm.
}
\end{figure*}

The second way, somewhat resembling the process described by Goldfarb {et al.}\cite{Nucleation} for the case of the gas-source MBE (and thick hydrogenated WL),  is illustrated by Fig.~\ref{fig:heap}. At small values of $h_{\rm Ge}$, regions containing excess of Ge atoms are observed on the surface. Usually, they  are not resolved as structured formations and resemble shapeless heaps of Ge (Fig.~\ref{fig:heap}a).  Terrace edges or pits\cite{footnote}
usually accompany them. Heap density is about $10^9$\,cm$^{-2}$. Some of heaps start to form the $\{$105$\}$ facets during Ge deposition (Fig.~\ref{fig:heap}b). 
Annealing at $T_{\rm gr}$ results in formation of volumetric structures partially  faceted by $\{$105$\}$ planes,    transforming heaps to some similarities of huts but with truncated very  disordered and rough apexes (Fig.~\ref{fig:heap}c--e). 
Formations of this type are also observed in Fig.~\ref{fig:nuclei600}c,d: they are adjacent to mature pyramids.

We have never  observed such formations and such process at low temperatures of growth and suppose them to be inherent only to the high-temperature assembly of huts. We assume that Ge heaps appearing at high temperatures are products of intense migration of Ge adatoms to surface irregularities, such as pits or terrace edges or sides of large pyramids, which are active attractors of highly mobile adatoms on Ge WL due to  WL local stress relaxation. Much less mobility (the hop rate)\cite{diffusion_semicond_surf} of adsorbed Ge atoms is likely a cause of absence of Ge heaps at low temperatures and formation of huts from the plane 16-dimer nuclei: huts nucleate before most of adatoms can reach sinks. Hut nucleation via formation of the 16-dimer embryos appears to be the most efficient sink of Ge adatoms at low temperatures, whereas at high temperatures, significant effect of competitive attractors redirect adatom fluxes and decrease hut nucleation rate.

An alternative explanation of the observed phenomenon may be as follows: Minor amount of carbon perhaps arriving to the surface and into the Ge film  from the residual atmosphere of the MBE vessel or from chemicals during Si substrate preparation, or even from  Czochralski-grown Si wafers  might also give rise to formation of the Ge heaps via Ge adatoms migration to WL areas with local strain lowered due to presence of carbon. A possibility of presence of some difficult-to-detect amount of carbon in the Ge films grown at 600 or 650{\textcelsius} is hinted by formation of the $c(4\times 4)$ reconstruction on the Si(001) surface during its dehydrogenation at the temperatures higher than 600{\textcelsius} in the MBE chamber after wet deoxidization of the Si surface in dilute aqueous solutions of the hydrofluoric acid. \cite{VCIAN2011} However, we have not got a verification of this assumption at present and believe that it is very unlike. Further experiments on Ge film growth on  buffer Si layers shall verify or refute this hypothesis.

\subsection{Reflected High-Energy Electron Diffraction}

\begin{figure*}
\includegraphics[scale=1]{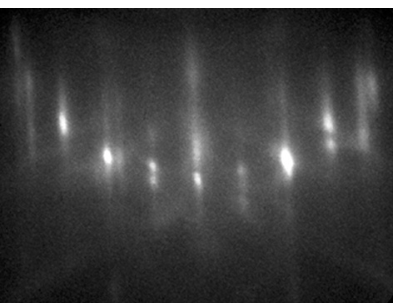}(a)
\includegraphics[scale=1]{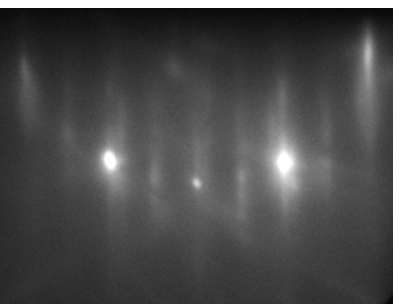}(b)\\
\includegraphics[scale=1]{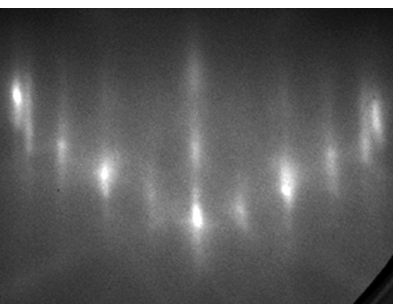}(c)
\includegraphics[scale=1]{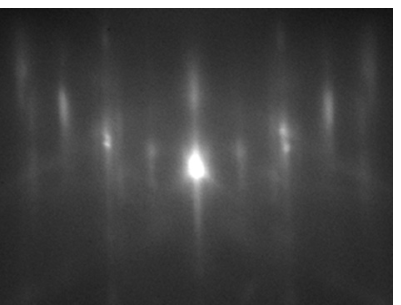}(d)\\
\includegraphics[scale=1]{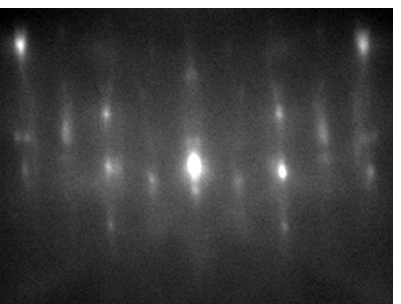}(e)~~~~~~~~~~~~~~~~~~~~~~~~~~~~~~~~~~~~
\caption{\label{fig:rheed}{\it In situ}
RHEED patterns of Ge/Si(001) films, $E = 10$\,keV, [110] azimuth:
(a)  $T_{\rm gr} =$ 650{\textcelsius}, $h_{\rm Ge}=$ 4\,\AA;
(b) $T_{\rm gr} =$ 360{\textcelsius}, $h_{\rm Ge}=$ 4\,\AA;
(c) $T_{\rm gr} =$ 650{\textcelsius}, $h_{\rm Ge}=$ 5\,\AA;
(d) $T_{\rm gr} =$ 650{\textcelsius}, $h_{\rm Ge}=$ 5\,\AA, annealing  at the deposition temperature for 7 min;
(e) $T_{\rm gr} =$ 650{\textcelsius}, $h_{\rm Ge}=$ 6\,\AA, the similar pattern is obtained for $T_{\rm gr} =$ 600{\textcelsius}.
}
\end{figure*}

 Diffraction patterns of reflected high-energy electrons for samples of thin ($h_{\rm Ge}=$ 4\,\AA) Ge/Si(001) films deposited at high (650 or 600\textcelsius) and low (360\textcelsius) temperatures with equal effective thicknesses   are presented in Fig.~\ref{fig:rheed}a,b. The patterns are similar and represent a typical $(2\times 1)$ structure of Ge WL; reflexes associated with appearance of huts (the 3D-reflexes) are absent in both images, that agrees with the data of the STM analysis. Diffraction patterns presented in Fig.~\ref{fig:rheed}a,c,e correspond to the samples with $h_{\rm Ge}$ increasing from 4 to 6\,\r{A}. The 3D-reflexes are observed only in the pattern of the samples with $h_{\rm Ge}=$ 6\,{\AA} , that is also in good agreement with the STM data.

Influence of the sample annealing at the deposition temperature is illustrated by a complimentary pair of the RHEED patterns given in Fig.~\ref{fig:rheed}c,d. Annealing of specimens at the temperature of growth resulted in appearance of the 3D-reflexes (Fig.~\ref{fig:rheed}d) that also corresponds to the results of our STM study.

Difference in dynamics of diffraction patterns during the deposition of Ge atoms is a characteristic feature of the high-temperature mode of growth in comparison with the low-temperature regime. The original Si(001) surface is $(2\times 1)$ reconstructed. At high temperatures, as $h_{\rm Ge}$ increases, diffraction patterns evolve as $(2\times 1)\rightarrow (1\times 1)\rightarrow (2\times 1)$ with very weak {\textonehalf}-reflexes. Brightness of the {\textonehalf}-reflexes increases (the $(2\times 1)$ structure becomes pronounced) and the 3D-reflexes arise only during sample cooling. At low temperatures, the structure changes as $(2\times 1)\rightarrow (1\times 1)\rightarrow (2\times 1)\rightarrow (2\times 1)+3$D-reflexes. This observation reflects the process of Ge cluster `condensation' from the 2D gas of mobile Ge adatoms. High Ge mobility and low cluster nucleation rate in comparison with fluxes to competitive  sinks of adatoms determines the observed difference in the surface structure formation at high temperatures as compared with that at low temperatures.

\section{Conclusion}

Summarizing the above we can conclude that, as distinct from the case of low-temperature deposition of Ge films on Si(001) when the only process of Ge hut appearance exists, two process resulting in emergence of $\{105\}$-faceted clusters on  Ge WL have been observed at high temperatures: Pyramids have been observed to nucleate via the previously described formation of strictly determined structures, resembling blossoms, composed by 16 dimers grouped  in pairs and chains of 4 dimes on tops of the wetting layer $M\times N$ patches, each on top of a separate single patch, just like it goes on at low temperatures; an alternative process consists in faceting of shapeless heaps of excess Ge atoms which arise in the vicinity of strong sinks of adatoms, such as pits or steps. The latter process has never been observed at low temperatures; it is typical only for the high-temperature deposition mode.

Different dynamics of RHEED patterns during the deposition of Ge atoms in different growth modes is observed, which reflects the difference in mobilities  and adatom `condensation' fluxes from Ge 2D  gas on the surface for different modes, which in turn  control the nucleation rates and densities of Ge clusters. 
  

\begin{acknowledgments}

 This research was supported by the Ministry of Education and Science of the Russian Federation under the State Contracts No.\,14.740.11.0069 and  No.\,16.513.11.3046.

\end{acknowledgments}


\begin{thebibliography}{28}
\bibitem [{foo()}]{footnote}%
  \BibitemOpen
  \href@noop {} {}\bibinfo {note} {The pits are different from the
  $\{105\}$-faceted ones observed in the gas-source-MBE grown Ge films and
  demonstrated in Ref.\,\onlinecite{Nucleation}}\BibitemShut {NoStop}%
\end{thebibliography}





%

\end{document}